\documentclass{emulateapj}

\usepackage{graphicx}

\usepackage{natbib} 

\newcommand{\cgs}{ erg cm$^{-2}$ s$^{-1}$ }

\newcommand{\source}{GS~2023+338}
\newcommand{\targ}{\object{V404 Cygni}}

\newcommand{\intg}{{\em INTEGRAL}}

\slugcomment{Accepted for publication in Astrophysical Journal Letters}

\shorttitle{X-ray spectral evolution of V404 Cygni in the initial phase of the 2015 outburst }
\shortauthors{Natalucci et al.}

\begin{document}

\title{High energy spectral evolution of {\emph V404 Cygni} during the June 2015 outburst as 
observed by \intg\/\altaffilmark{1}}
 
\altaffiltext{1} {\intg\/ is an ESA project with instruments 
 and science data centre funded by ESA member states 
(especially the PI countries: Denmark, France, Germany, Italy, Switzerland, Spain),
 Czech Republic and Poland, and with the participation of Russia and the USA.}

\author{Lorenzo Natalucci, Mariateresa Fiocchi, Angela Bazzano \& Pietro Ubertini}
\affil{Istituto di Astrofisica e Planetologia Spaziali, INAF, Via Fosso del Cavaliere 100, Roma, I-00133, Italy} 
\author{Jean-Pierre Roques \& Elisabeth Jourdain}
\affil{Universit\'e de Toulouse; UPS-OMP\\ CNRS; IRAP; 9 Av. Colonel Roche, BP 44346, 
F-31028 Toulouse cedex 4, France} 
\begin{abstract}
The black hole binary \source\/ exhibited an unprecedently bright outburst on June 2015. Since June 17th, 
the high energy 
instruments on board \intg\/ detected an extremely variable emission during both bright and low luminosity phases, 
with dramatic variations of the hardness ratio on time scales of $\sim$seconds. The analysis of the IBIS and SPI data 
reveals the presence of hard spectra in the brightest phases, compatible with 
thermal Comptonization with temperature $kT_e\sim40$~keV. The seed photons temperature is best fit by 
$kT_0\sim7$~keV, that is too high to be compatible with blackbody emission from the disk. 
This result is  
consistent with the seed photons being provided by a different source, that we hypothesize to be a synchrotron 
driven component in the jet.       
During the brightest phase of flares, the hardness shows a complex pattern of correlation with flux, with a maximum 
energy released in the range $40-100$~keV. The hard X-ray variability for $E>50$~keV is correlated with flux variations 
in the softer band, showing that the overall source variability cannot originate entirely from absorption, but 
at least part of it is due to the central accreting source.
\end{abstract}
\section{Introduction}
The recurrent black hole binary \source(=\targ) went into outburst on June 15th, 2015
after 26 years of quietness. A sudden increase of accretion rate was initially detected with 
the {\em Swift} satellite \citep{bar15} and reported also by {\em MAXI} \citep{neg15}. Soon after these notifications, 
many ground and space facilities have been following the event. 
Thanks to its exceptionally bright flux, \targ~ 
represents an excellent test laboratory at high energies, where it is possible to unveal the properties of the central 
engine thanks to the reduced absorption effect. 
\targ\/ was  detected for the first time in X-ray  
with the {\em Ginga} satellite in 1989, 
May 19 \citep{mak89,kit89} and then observed with instruments on board the 
{\em Roentgen} observatory 
on Mir-Kvant for more than 2 months \citep{sun91}. 
The authors reported on the emission spectra in the 2-300 KeV band at 
different epochs and different flux levels, with evidence of variability of the soft part (E$<$15 keV). 
The source behaviour was also followed up in optical and the system identified with \targ\/ \citep{wag89}, 
while use of archival optical data allowed the discovery of two outbursts in 1938 and 1956. The extensive monitoring 
of the source allowed the determination of the key parameters of the system, having a mass function of 
6.08$\pm$0.06 \citep{cas94}. The compact object is a $\sim10{M}_{\rm \odot}$ black hole with an orbital 
period of 6.5 days and a $1{M}_{\rm \odot}$ K0IV companion \citep{cas93}. 
The best distance estimate is $2.39\pm0.14$~kpc \citep{mil08}. 
This source has been the best studied quiescent system being also the most luminous stellar black hole in this 
state \citep{bra07,hyn09,ran15}. A spectral energy distribution  with simultaneous 
coverage at X, UV, optical and radio has been compiled, revealing an IR excess, originated from either a cold disk 
or a jet \citep{hyn09}. The X-ray data from two epochs indicated a power law continuum with index $\sim2$, with no indication of 
spectral variability in spite of a factor 10 variation 
in luminosity. The radio data is consistent with a flat spectrum, resembling the typical X-ray binaries hard state and 
associated with a compact jet; the radio variability is not correlated with X-ray variability.
 
\begin{figure}
   \includegraphics[angle=-90,width=8.5cm]{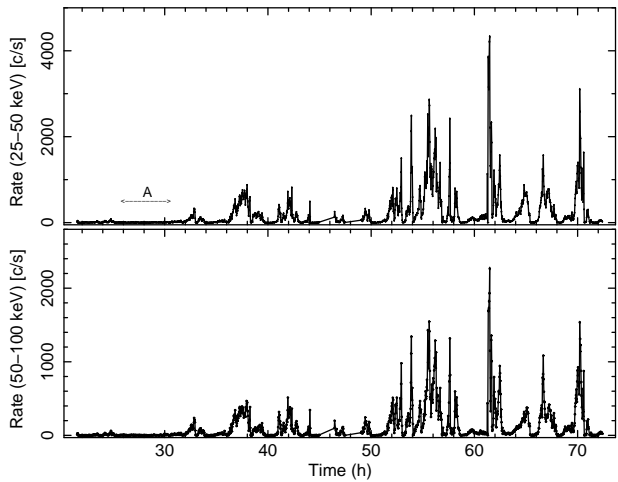}
   \caption{Light curve with IBIS/ISGRI count rates in the 25-50 keV and 50-100 keV energy bands, for the data period
spanning revolution 1554. Start time is MJD~57190, UTC 21:30:45:186.}
   \label{fig1}
   \end{figure}
 
On June 17th, 2015 a public \intg\/ Target of Opportunity (ToO) program started and we here report on the 
preliminary analysis of the collected data.

\begin{figure}
   \includegraphics[angle=-90,width=8.8cm]{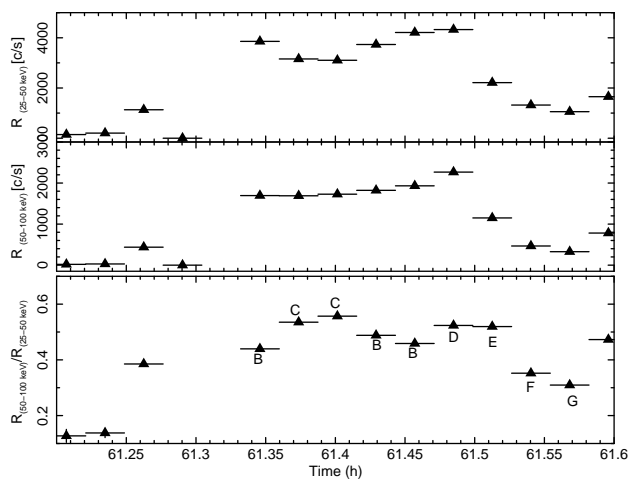}
   \caption{Light curves of the major flare in the 25-50 keV (top panel) and 50-100 keV (center panel) bands. 
Start time is MJD~57190, 21:30:45:186 UT. The 
hardness ratio in the two bands is shown in the bottom panel. Periods B to G mark the intervals used for the
spectral analysis (see text for details).}
   \label{fig2}
   \end{figure}
\section{Observations and data analysis} \label{sect_obs}
V404 Cyg has been monitored at high energies by \intg\/ and other space facilities by dedicated ToO programs 
\citep{fer15a,roq15a,rod15a,kuu15,nat15,mot15a,mot15b,seg15,gan15}.
An initial, extended phase with intense flaring and large, 
short term variations in hardness (down to a few seconds time scales) occurred, with the source reaching 
luminosities up $\sim5\times10^{38}$~erg~s$^{-1}$.
Most instruments sensitive in the X-ray band were saturated by such a high flux, see e.g. \citet{kin15a}. 
The peak intensity during ~5 days of activity reached 44 Crab on MJD 57194.31 as detected by \intg\/ 
\citep{rod15b}. 
\intg\/ monitored the source quite continuously since June 17th, with short ($\sim$~half a day) 
interruptions due to operational 
constraints (mostly, scheduled for each perigee passage). Around June 28th, the source had much 
faded \citep{fer15b} but remained visible at a level of a few mCrab intensity in the hard X-rays \citep{wal15}. 

In this paper we analyze the spectral evolution of \targ~ during the intense initial flaring. We use \intg\/ observations 
performed during revolution 1554, starting at MJD~57190.896 with a total elapsed time of 50.55 hours. 
This period is coincident with the starting phase of the \intg\/ monitoring. 
The IBIS/ISGRI data for this observation are near real-time data processed using the latest release of the \intg\/ 
Offline Scientific Analysis (OSA~V10.1). This software has been used to obtain light curves and spectra in 
different common time intervals. SPI data are taken from the consolidated data archive, whereas their 
analysis makes use of specific tools developed at IRAP \citep{roq15b}.

The time evolution of the hard X-ray emission from the source is shown in Figure~1. 
The hardness ratio (hereafter, HR) during the most turbulent phases varies on time scales down to at 
least a few seconds. In Figure 2 the intensity and hardness profiles are shown for the 
science window (hereafter, SCW) no.43, where the source flux reached its maximum within our observing period. 
For the purpose of spectral analysis, we divided the time profile of the major flare into time 
intervals 100s long, as shown in Figure~2, and built a total of six spectra by grouping the ones with    
similar HR and fluxes. In particular, we group spectra if their flux difference 
is $<20$\% and the difference in HR is $<0.05$. 
Together with the low state spectrum of period A this results in a set of seven spectra, 
A to G. For IBIS/ISGRI, the range 13-1000~keV 
was sampled using 85 channels with variable width, chosen according to a logarithmic law. We restricted the analysis 
to the range 25-300 keV, for which there exists an accurate response. 
\section {Results}  \label{sect_analysis}
The overall dependence of source hardness on hard X-ray flux is shown in Figure~3. 
The distribution of values, when represented as the (normalized) difference of count rates in the 
50-100 and 25-50 keV bands, reproduces the shape of a horse head. It is interesting 
to note that, while at low fluxes (below $\sim0.1$~Crab) the hardness is varying on a wide scale, 
it is more stable at intermediate 
fluxes with a weak trend of being correlated with flux up to intensities below $\sim2$~Crab. 
At higher fluxes, corresponding to the brightest flaring episodes, the hardness seems to be anticorrelated 
with flux. However, within a single X-ray flare, harder
spectra may occur at higher fluxes, as it holds for the green points corresponding to the data for 
the brightest flare, 
occurred within SCW 43 (intervals B to G). If this is the case, the anticorrelation HR-flux 
could hold for peak fluxes of
different flare episodes and not during a single flare (see also Figure~4).    

The IBIS/ISGRI data are affected by gaps due to limited telemetry bandwidth. The length of such gaps 
is of the order of a few seconds and their rate is slightly variable, being correlated with source intensity; due to this, the software actually reconstructs only
a limited portion of spectral information for some of the 100s long spectra.
For this reason we have been able to obtain a good match for IBIS and SPI data (i.e. same live time intervals) 
only for the periods A, F and G.

 \begin{figure}[h]
   \includegraphics[angle=-90,width=8.7cm]{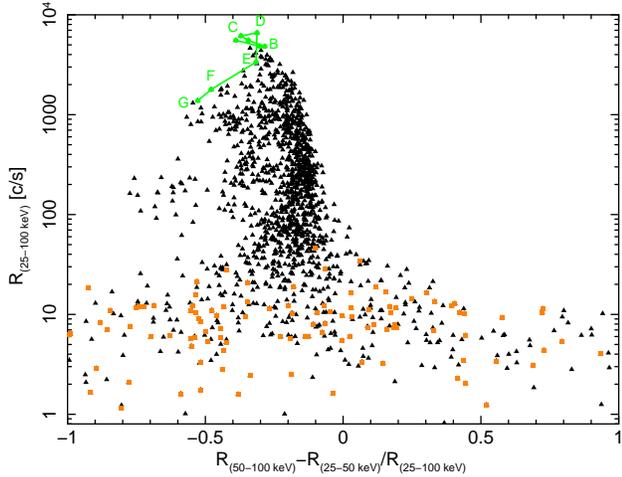}
   \caption{Intensity vs. hardness diagram for the full data set. Each point is generated using intensity values 
averaged over 100s. Square orange points at the bottom are integrated on period A, whereas green circle points 
(B to G) are taken during the most intense flare. }
   \label{fig3}
   \end{figure}

The spectra shown in Figure~4 describe the evolution of the source emission during the low state period (A) 
and the flare of SCW~43 (B to G), as derived with IBIS/ISGRI and SPI. IBIS and SPI source fluxes have been generally 
found in good agreement ($\leqslant12\%$), despite the strong source variability and the non-homogeneous data taking 
intervals for the two instruments. 
Furthermore we found that, at high flux levels, the IBIS/ISGRI spectra appear systematically harder than SPI ones,
yielding a difference in spectral index lower than $\sim5-10$\%.  
The reasons for this are under investigation; however, we have verified that the induced systematic effect on 
the spectral parameters is very small when fitting with physically motivated models.

The low state spectrum measured for period A (see Figure~1) has been analyzed by performing a joint fit with
both instruments. It can be modeled 
by a simple power law with photon index $\Gamma=1.5\pm0.3$, with a lower limit for the high energy cutoff 
of $\approx120$~keV. The average flux is $1.8\pm0.1\times10^{-9}$\cgs in the 25-400~keV band, 
corresponding to a luminosity of $\sim1.2\times10^{36}$ erg s$^{-1}$. The two instruments cross-normalization 
is $1.0\pm0.1$. 

The spectra during the main 
flare (B to G) show a prominent high energy cutoff. A thermal Comptonization model, {\em comptt} in XSPEC, 
provides a satisfactory description of the data (see Table~1). For the highest fluxes, B to E, the temperature of the 
accreting plasma is quite constant around $\sim40$~keV. After some flux decline (spectrum F)
the electron temperature increases up to $>100$~keV and the optical depth decreases ($\tau<0.1$). In all cases, 
the temperature of the seed photons for the Comptonization is high, around $\sim7$~keV. We have verified that a
model spectrum with a seed photon temperature of between 0.1 and 2~keV, typical of the thermal blackbody emission 
from an accretion disk, cannot reproduce the data within the thermal Comptonization model. Forcing the
seed temperature to such low values, the fit requires a second additive component, which
cannot be accounted by reflection: in this case,    
modeling the spectra with a single direct continuum with reflection yields excessively high 
values of the reflection normalization, as we verified using   
both the {\em pexriv/pexrav} \citep{mag95} and {\em xillver} \citep{gar13} models.
Conversely, if we adopt the single component model reported in Table~1, the use of reflection does not
improve the fit results, with the exception of spectrum C, for which fitting the data with an added {\em xillver} 
component yields a significantly better result ($\chi^{2}_{\rm red}=1.12$ instead of 1.41).  
In this case, the fraction of reflected flux is found to be $\approx22$\%.

Concerning our detection of the $\sim7$~keV soft component, we considered effects that 
could modify the shape of the spectrum at the lowest energies, like partial covering of the 
central X-ray source.
\citet{kin15b} reported about two soft X-ray observations with {\em Chandra} taken on June 22 and 23, 
when \targ~ was still active with average 2-10~keV fluxes of $9.5\times10^{-9}$\cgs and 
$1.3\times10^{-8}$\cgs, respectively. The detection of strong narrow emission lines led the authors to
conclude that the central source is obscured, with evidence for the line formation region to be located
at distances greater than $7\times10^{-9}$~cm, possibly coincident with the outer disk. 
Motivated by this result we used a partial covering absorber ({\em pcfabs}) to model our flare spectra (B to G), 
in conjunction with the usual {\em comptt} component. As expected, we obtain very large errors in the 
parameters characterizing the low energy spectrum, i.e. the covering fraction
$F_{cov}$, the related absorption column, $N_{h}$, and the seed temperature $kT_{0}$. 
Nevertheless, if we impose values of $kT_{0}$ lower than 2 keV in these fits, the $N_{h}$ values are always 
far too high, i.e. typically between $5\times10^{24}$~cm$^{-2}$ and $3\times10^{25}$~cm$^{-2}$ 
for values of $F_{cov}$ from 0.20 to unity. 
We conclude that the partial covering of the source does not affect our main result about the observed high 
temperature of the seed photon spectrum.  

Finally, fitting the IBIS/ISGRI data with the hybrid thermal/non-thermal Comptonization model {\em eqpair} 
\citep{cop99} also yields good results: $\chi^{2}$/dof=48.6/37 and 25.9/37 for spectrum~C
and spectrum F, respectively. Spectrum~C requires high values of the reflection 
normalization ($\Omega/2\pi\sim1$). 
The temperature of
the seed photon distribution for the above fits is $kT_{bb}=6.2_{-0.4}^{+0.9}$ and 
$kT_{bb}=7.6_{-0.9}^{+0.5}$ respectively, essentially in good agreement with the {\em comptt} model.    

\begin{table}
\begin{center}
\scriptsize
\caption{Results of spectra fitting using the {\em comptt} model.}
\label{fit}
\begin{tabular}{lccccr}
\tableline\\ [-2.0ex]
Spec.Id$^a$ & kT$_0 (keV)$ & kT$_e (keV)$ &  $\tau$ & Flux$^b$ &  $\chi^{2}$/dof \\
\tableline\\ [-2.0ex]
B & $7.05\pm0.2$    & $42_{-4}^{+7}$ & $0.7_{-0.2}^{+0.1}$ &  $5.5$ & 25.8/38 \\
C & $7.4\pm0.2$ & $40_{-3}^{+5}$ & $1.0\pm0.16$  & $5.0$  & 54.9/39 \\
C$_{R}^c$ & $8.6_{-1.4}^{+1.0}$ & $42_{-18}^{+8}$ & $0.9\pm0.3$ & $5.0$ & 40.3/36 \\
D & $7.2\pm0.3$ & $34_{-2}^{+3}$ & $1.2\pm0.2$ & $6.6$ & 42.0/37 \\
E & $7.4\pm0.3$ & $41_{-5}^{+11}$ & $0.9_{-0.3}^{+0.2}$ & $3.5$ & 37.9/38 \\
F$^{(*)}$ & $6.4\pm0.2$ & $182_{-93}^{+8}$ & $<0.09$ & $1.73^d$ & 91.5/73 \\
         &   & & & $1.76^e$ & \\
G$^{(*)}$ & $6.0\pm0.3$ & $63_{-33}^{+88}$ & $<0.7$ & $1.17^d$ & 101.3/66 \\
         &   & & & $1.04^e$ & \\
\tableline\\ [-2.0ex]
\end{tabular}
\tablecomments{
$^a$ Fits with IBIS and SPI spectra are marked by $^{(*)}$. In all other cases, 
only IBIS data have been used.  
$^b$ Flux values in units of $10^{-7}$erg~s$^{-1}$~cm$^{-2}$, in the 20-200~keV range.
$^c$ Fit with added reflection component (see text for more details). 
$^d$ Flux measured by IBIS.
$^e$ Flux measured by SPI. }
\end{center}
\end{table}
\section{Discussion} \label{sect_disc}
\targ\/ is an established transient black hole binary, for which two bright outbursts 
have been available for study in the hard X-rays. On the basis of its behaviour in the previous 1989 
outburst, it had been 
proposed \citep{tom04} to belong to a class of transient sources for which the spectral state is 
quite independent on luminosity. 
Figure~3 confirms this picture as a general trend also for this most recent event, at least for source 
luminosities above $\sim10^{36}$~erg~s$^{-1}$. 

\begin{figure}[ht]
\begin{center}
   \includegraphics[angle=-90,width=8.7cm]{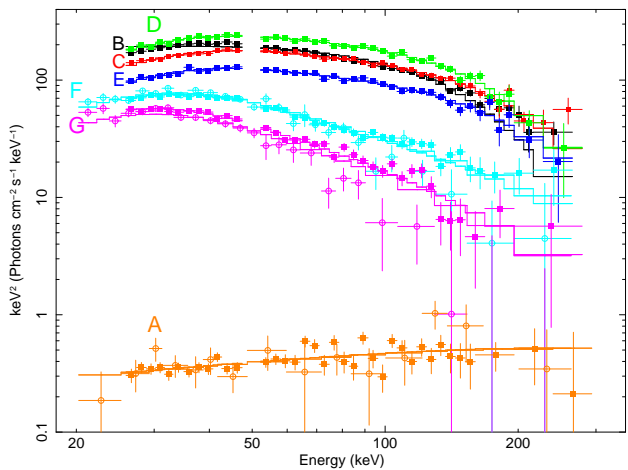}
   \caption{Spectra measured for different phases: a low state spectrum (A) and spectra  
measured during the most intense flare (B to G). Circles denote data from SPI; squares from IBIS/ISGRI.}
   \label{fig4}
   \end{center}
   \end{figure}

One of the questions concerns the origin of the X-ray variability, i.e. is this related to rapid changes in the  
accretion flow, to a variable component from a jet, or both?  
From the hard X-ray data we can exclude that the variability is totally due to absorption by a disk outflow or wind. 
Whereas in the soft X-rays the central source could be partially or totally obscured \citep{kin15b}, using the hard X-rays 
we are unveiling the variability component of the central accreting region, as proven by the 
rapid and large intensity fluctuations detected at energies up to $>100$~keV. 

During the current outburst of \targ, we observed that the spectrum of the main flare is actually described 
by Comptonization of a thermal plasma. The size of the corona can be estimated following the approach of
\citet{mer01} using our measured parameters of temperature, accretion rate, luminosity and optical depth. 
If the energy of the corona is purely thermal, its size would be of the order
of $\sim10^{4}$ Schwarzchild radii ($R_s$). Conversely, if the stored energy is mainly present as   
a magnetic field the size would be of the order of a few $R_s$. Such a compact corona could explain,
in our case, the almost constant plasma temperature of $\sim40$~keV which is observed in all high flux spectra
of the main flare.
This scenario could also favor the possibility that the seed photons are not injected by the 
accretion disk, but have different origin as imposed by their high temperature of $\sim7$~keV.
The seed photons could be provided by a moderately optically thick synchrotron source within the jet, 
giving rise to a self-absorbed spectrum. This radiation would be then upscattered by the surrounding 
hot corona. For reference, a scenario with synchrotron generated seed photons has been discussed
by \citet{war00}.  

Finally, we note that our spectral results are not affected by  
instrumental effects due to the exceptionally high count rate from \targ. 
In fact, they are in agreement with the spectral analysis reported 
by \citet{roq15b} using the SPI data for the same observations. The SPI and IBIS/ISGRI detectors are based on 
completely different technology and are cross-calibrated in-flight using the Crab Nebula as 
reference source. 
\section{Conclusions} \label{sect_concl}
We have analyzed \intg\/ data from the initial phase (first pointing with \intg) of the \targ\/ recent outburst. 
The source features high flux variability and hard spectra 
even in the brightest flares, with evidence for a thermally Comptonized spectrum 
with a plasma temperature of $\sim40$~keV. Despite
the spectral parameters of the analysis must be taken with caution due the intrinsic limitations of the 
models, there is indication for a seed photon component not coincident with the hot accretion disk 
thermal photons. 

By means of this study we characterize the high energy behaviour of the source at the highest fluxes,
where the hardness during single flares is correlated with flux (see Figure 4) up to luminosities of a few 
percent of Eddington, and anti-correlated at higher fluxes. The general behaviour in hard
X-rays is in agreement 
with the spectral variability of the source detected in the previous 1989 outburst (Sunyaev et al. 1991). 
\acknowledgments
The Italian authors acknowledge ASI/INAF agreement n. 2013-025-R.0 and thank M.~Federici 
for setup and maintenance of the \intg~archive and analysis S/W at IAPS.  
The \intg/SPI project has been completed
under the responsibility and leadership of CNES. We are grateful to ASI, CEA, CNES, DLR, 
ESA, INTA, NASA and OSTC for support. We also thank the anonymous referee for her/his precious
suggestions.

\end{document}